\documentstyle[natbib,amssymb,graphicx]{aipproc}

\newlength{\figurewidth}
\begin{document}

\title{RXTE Monitoring of LMC~X-3: Recurrent Hard States}
\author{J.~Wilms$^*$, M.A.~Nowak$^\dagger$,
  K.~Pottschmidt$^*$, W.A.~Heindl$^\ddagger$,
  J.B.~Dove$^{\|,\P}$, M.C.~Begelman$^{\dagger,\S}$, 
  R.~Staubert$^*$}

\address{$^*$ Institut f\"ur Astronomie und Astrophysik, Waldh\"auser Str. 64, D-72076 T\"ubingen, Germany\\
  $^\dagger$ {JILA, University of Colorado, Boulder, CO
    80309-440, U.S.A.}\\
  $^\ddagger${CASS, University of California, San Diego, La
    Jolla, CA 92093, U.S.A.}\\
  $^\|${CASA, University of Colorado, Boulder, CO
    80309-389, U.S.A.}\\
  $^\P${Dept.\ of Physics, Metropolitan State College of
    Denver,  Denver, CO 80217-3362, U.S.A.}\\
  $^\S${APS, University of Colorado, Boulder 80309, U.S.A.}}

\maketitle

\begin{abstract}
  The black hole candidate LMC X-3 varies by a factor of four in the soft
  X-rays on a
  timescale of  200 or 100\,days \citep{cowley:91a}.  We have
  monitored LMC X-3 with RXTE in three to four week intervals starting in
  December 1996, obtaining a large observational database that sheds light
  on the nature of the long term X-ray variability in this source. In this
  paper we present the results of this monitoring campaign, focusing on
  evidence of recurring hard states in this canonical soft state black hole
  candidate.
\end{abstract}

\section*{Introduction}
Long-term X-ray variability on timescales of months to years is seen in
many galactic black hole candidates. In analogy to the 35\,d cycle of
\mbox{Her~X-1}, this variability has been identified with the precession of
a warped accretion disk in some objects.  Possible driving mechanisms
include radiation pressure \citep{maloney:98a,wijers:99a} or accretion disk
winds \citep{schandl:96a}.  In this paper we present a spectral and
temporal analysis of the long term variability of the canonical soft state
black hole candidate LMC~X-3. Together with LMC~X-1, this source is the
only persistent black hole candidate which so far has only been observed in
the soft state. While LMC~X-1 does not exhibit any long term variability,
LMC~X-3 was known to be variable on a $\sim$100\,d timescale
\citep{cowley:91a,cowley:94a}. Detailed results of our campaign are
presented elsewhere \citep{wilms:99d}.

\section*{Long Term Variability}

Our analysis of the long-term RXTE All Sky Monitor light curve (Fig.
\ref{fig:x3temp}, top) indicates a complex long term behavior. Analysis
with the \cite{lomb:76a}-\cite{scargle:82a} Periodogram indicates that the
variation is dominated by epochs of low luminosity, which are recurring on
the $\sim$100\,d timescale found previously \citep{cowley:94a}. In
addition, a long term periodicity is apparent in the data. Contrary to the
100\,d timescale, the long term periodicity is not stable: Depending on
what time interval of the ASM light curve is studied, the long term period
varies between 200 and 300\,d. This periodicity is caused by the times of
average to high luminosity seen in the light curve and manifests itself by
a broad peak at $\sim 250$\,d in the Lomb Scargle PSD.

\section*{Spectral Variability}

We have analyzed the RXTE data using the newest RXTE \texttt{ftools}, as
well as \texttt{XSPEC}, Version~10.00ab. The spectral model used for the
data analysis was the standard multi-temperature disk blackbody
\citep{mitsuda:84a,makishima:86a}, plus a power-law component.  Adding a
Gaussian iron line resulted in upper limits for the line equivalent width
only. Typical reduced $\chi^2$ values were $\chi^2_{\rm red}<2.5$ for 41
degrees of freedom, with the residuals being fully consistent with the
uncertainty of the detector calibration \citep{wilms:98c,wilms:99d}.

\begin{figure}
\centerline{\includegraphics[width=0.9\textwidth]{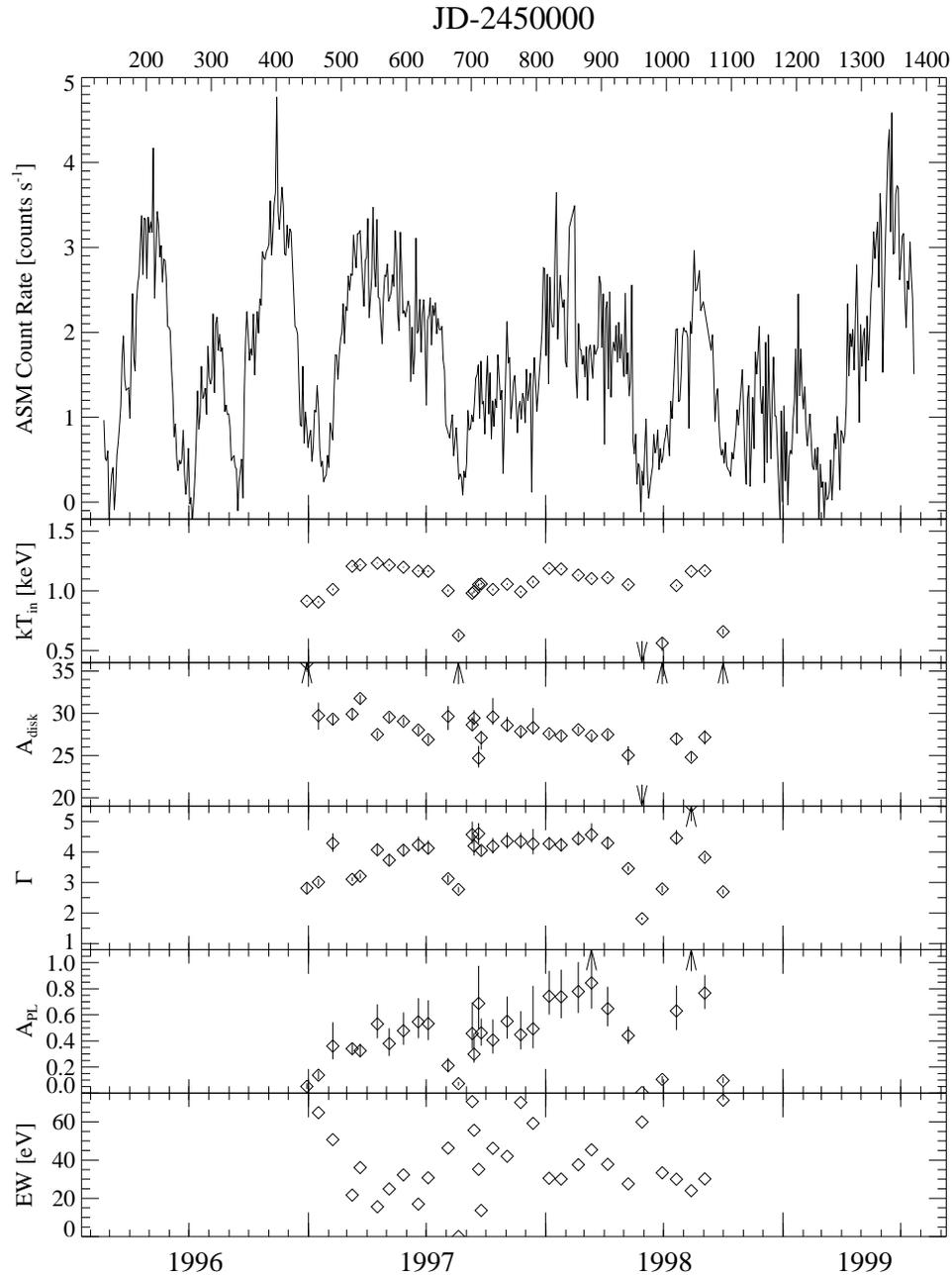}}
\caption{Temporal variability of the spectral parameters of
LMC~X-3. Note that the power law is harder and the inner disk
temperature is smaller during the recurring episodes of low
source luminosity.}\label{fig:x3temp} 
\end{figure}

In Fig. \ref{fig:x3temp} we present the variation of the spectral
parameters found during the analysis as a function of time. During episodes
of high ASM flux, the source behaves like any other source in the classical
soft state: The accretion disk temperature, $kT_{\rm in}$ varies freely to
accommodate the variable luminosity of the source, while the the
normalization of the multi-temperature disk black body is constant.  At the
same time, the photon index $\Gamma$ varies independently of $kT_{\rm in}$.
See, e.g., \citet{tanaka:95a} for similar examples in other soft state
black hole candidates.

On the other hand, for times of low ASM count rate, the disk temperature
decreases to $kT_{\rm in} \ll 1$\,keV from its usual value of $\sim
1$\,keV, while at the same time the photon index changes dramatically from
$\sim 4$ to $\sim 1.7$. We interpret these changes as evidence for
transitions to the hard state in LMC X-3.

\section*{Hard States in LMC~X-3}

\begin{figure}
\centerline{\includegraphics[width=0.6\textwidth]{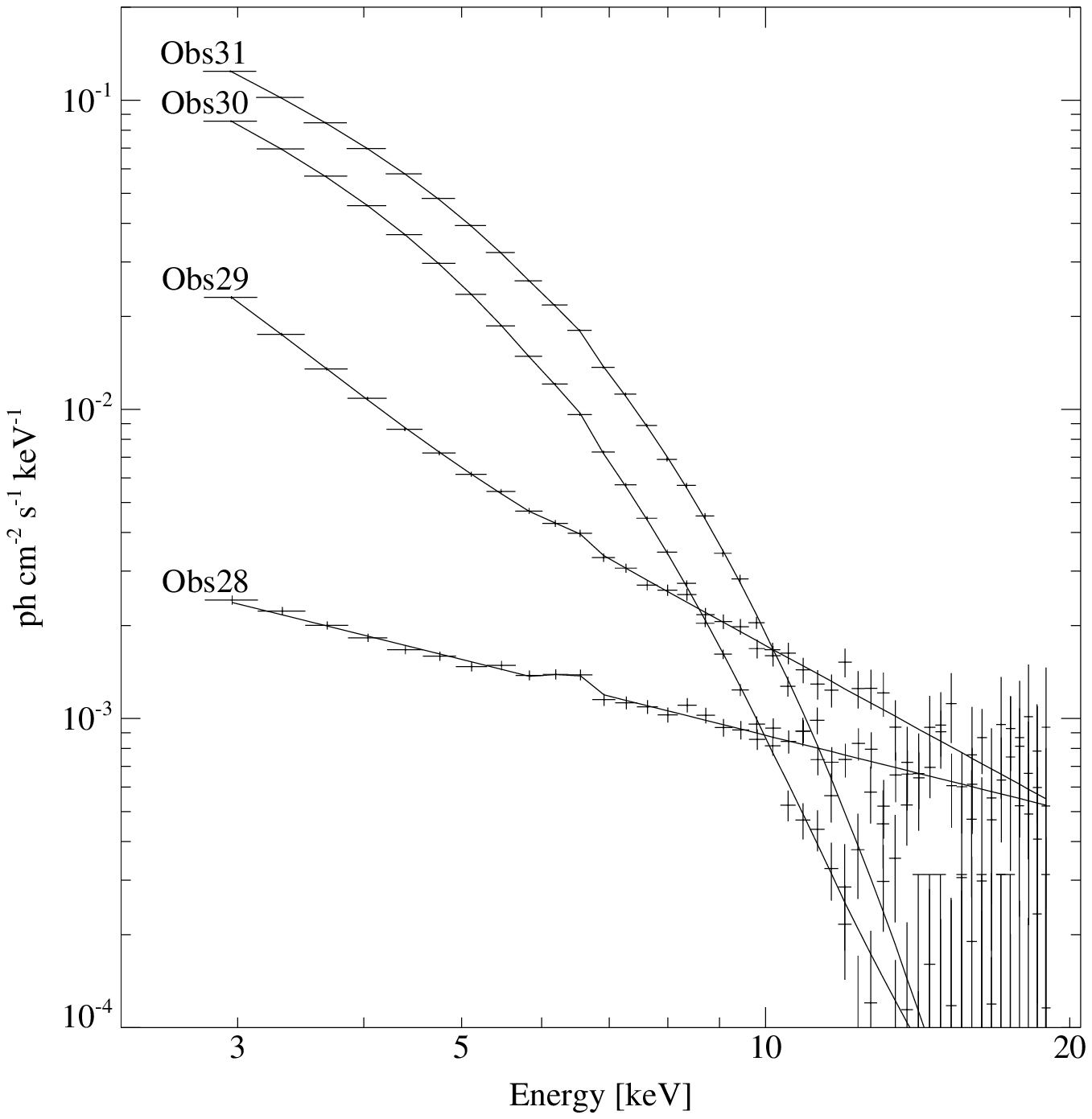}}
\caption{Spectral evolution from the hard state of 1998 May 29
  (JD=2450962; Obs28) to the normal soft state spectrum in 1998 July and
  August (Obs30 and Obs31). Shown are the unfolded photon spectra, the
  lines denote the best fit model.}\label{fig:x3specvar}
\end{figure}

Fig.~\ref{fig:x3specvar} displays the spectral evolution of
LMC~X-3 from 1998~June through August. In Obs28 the source had
the lowest flux of all monitoring observations. No evidence for a
soft spectral component is present in the data, the spectrum is
consistent with a pure power-law spectrum with photon-index 1.8. After 
Obs28, the soft component slowly emerged until the standard
soft state spectrum is reached. 

\section*{Discussion and Conclusions}
We have presented the results from the first two years of our
RXTE campaign on LMC~X-3. This is the first campaign where a
systematic study of a soft state black hole candidate with
monthly coverage was possible \citep[earlier campaigns, such as
those of][suffered from the inflexible pointing constraints
of the earlier satellites]{cowley:91a,ebisawa:93a}. We have found
that the long-term luminosity variations are due to changes in
the spectral shape of the source, for large luminosity these
changes are due to a variation of the characteristic disk
temperature, $kT_{\rm in}$, while for small luminosity the source
undergoes a spectral hardening.  We have presented the first
clear case for a soft to hard state transition in this canonical
soft state black hole candidate.

Our results are a challenge to models in which the long term
variability of sources such as LMC~X-3 is explained in the
context of warped accretion disk models. In these models,
no clear spectral evolution with source intensity is expected,
with the exception of possible changes in $N_{\rm H}$ due to
covering effects. In black hole candidates such as Cyg~X-1, the
hard- to soft-state transitions are attributed to changes in the
accretion disk geometry, e.g., the (non-) existence of a
hot and Comptonizing electron cloud in the center of the
source. These changes are typically attributed to a varying mass accretion
rate, $\dot{M}$. Our result makes such a geometry also probable for
LMC~X-3. A possible cause for the quasi-periodicity of the soft
to hard transitions, therefore, might be periodic changes in $\dot{M}$.

\subsection*{Acknowledgements}
We thank the RXTE schedulers for their patience in scheduling a total of up
to now 40\,observations on LMC~X-3 and 32 observations on LMC~X-1 during
the course of the past three years. We also thank those who are responsible
for the almost non-existent observing constraints of RXTE for providing us
with the ability to perform campaigns such as this one. The attendance of
JW at the Compton symposium was made possible by a travel grant from the
Deutsche Forschungsgemeinschaft.

\parskip0pt
\bibsep0pt
\bibliographystyle{jwapjbib}
\bibliography{mnemonic,jw_abbrv,apj_abbrv,diplom,agn,ns,accret,bhc,inst,conferences}

\end{document}